\documentclass[fleqn,10pt]{wlscirep}
\usepackage[utf8]{inputenc}
\usepackage[T1]{fontenc}

\usepackage{graphicx} 

\usepackage{amsmath,amsthm,amssymb,mathrsfs}
\usepackage{bm}

\usepackage{color}

\title{Echo state property and memory capacity of artificial spin ice}

\author[*]{Tomohiro Taniguchi}
\affil[1]{National Institute of Advanced Industrial Science and Technology (AIST), Research Center for Emerging Computing Technologies, Tsukuba, Ibaraki 305-8568, Japan}

\affil[*]{tomohiro-taniguchi@aist.go.jp}

\begin{abstract}
Physical reservoir computing by using artificial spin ice (ASI) has been proposed on the basis of both numerical and experimental analyses. 
ASI is a many-body system consisting of ferromagnets with various interactions. 
Recently, fabricating magnetic tunnel junctions (MTJs) as ferromagnets in an ASI was achieved in the experiment, which enables an electrical detection of magnetic state of each MTJ independently.  
However, performing a recognition task of time-dependent signal by such an MTJ-based ASI has not been reported yet. 
In this work, we examine numerical simulation of a recognition task of time-dependent input and evaluate short-term memory and parity-check capacities. 
These capacities change significantly when the magnitude of the input magnetic field is comparable to a value around which the magnetization alignment is greatly affected by the dipole interaction. 
It implies that the presence of the dipole interaction results in a loss of echo state property. 
This point was clarified by evaluating Lyapunov exponent and confirming that the drastic change of the memory capacities appears near the boundary between negative and zero exponents, which corresponds to the edge of echo state property. 
\end{abstract}

\begin{document}

\flushbottom
\maketitle
%
%



Physical reservoir computing \cite{jaeger02,maas02,jaeger04,verstraeten07,appeltant11,dambre12,sillin13,tanaka19,nakajima21} is one type of information processing scheme utilizing nonlinear response from many-body systems as computational resource. 
It enables to recognize and/or predict time-series data, such as human voice and movie, and therefore, will be applicable for several purposes including natural language processing. 
Physical reservoir computing has been examined in systems having various physical systems such as soft matter \cite{nakajima14,nakajima15}, quantum matter \cite{fujii17}, and opitcal system \cite{fiers14,nakayama16,pauwels19,harkhoe19,brunner19}. 
In the field of spintronics, physical reservoir computing has been examined in various magnetic structure, such as spin-torque oscillator  \cite{torrejon17,furuta17,tsunegi18,riou19,kanao19,yamaguchi20,tsunegi23}, magnetic skyrmion \cite{bourianoff18,prychynenko18,raab22,lee23}, voltage-controlled memory \cite{taniguchi22} and spin wave \cite{nakane18,watt20,watt21,namiki23,namiki24,iihama24}. 
Each system has interesting features. 
For example, using spin-torque oscillator will enable all-electrical manipulation of device in nanometer scale, and thus, will be applicable to edge computing. 
A slow propagation speed of spin wave might contribute to a relatively long memory functionality. 


Among such ferromagnetic-based reservoirs, artificial spin ice (ASI) is another candidate that can be applied to physical reservoir computing. 
ASI is a many-body system consisting of ferromagnets and has frustration due to magnetic interactions such as local exchange interaction and/or dipole interaction \cite{bramwell01,wang06,tanaka06,qi08,gartside18,skjaervo20}. 
The presence of the frustration, or non-uniqueness of ground state, has a possibility to be used in distinguishing complex time-dependent data. 
Physical reservoir computing by ASI has been recently proposed in both numerical \cite{hon21} and experimental \cite{gartside22,hu23} studies, where the magnetic states were read by spin wave propagation or magnetic force microscopic image. 
In these ASIs, an external magnetic field is a necessary factor for computation as input signal. 
Recently, on the other hand, a honeycomb ASI consisting of magnetic tunnel junctions (MTJs) in nanometer scale \cite{kubota23} was experimentally fabricated \cite{kubota24}. 
This MTJ-based ASI has a feasibility in realizing all-electrically-controlled ASI, and thus, is suitable for several applications such as edge computing, although currently only the reading of the magnetic states of the MTJs was done electrically while its manipulation was still carried on by applying an external magnetic field \cite{kubota23}. 
However, a recognition task of such an MTJs-based ASI has not been examined yet. 


In this work, we perform numerical simulation of the Landau-Lifshitz-Gilbert (LLG) equation of an MTJ-based honeycomb ASI and evaluate its computational ability. 
Firstly, a saturation magnetization curve by applying an external magnetic field is evaluated. 
The curve shows non-monotonic behavior at a certain magnetic field strength, which originates from the dipole interactions between the MTJs. 
Secondly, the short-term memory and parity-check capacities are evaluated as a figure of merit of the computational ability. 
Here, the external magnetic field is used as input signal, and its strength and applied angle are widely changed. 
A drastic reduction of these capacities is observed near the magnetic field strength giving the non-monotonic behavior of the magnetization saturation. 
It implies that the dipole interaction between the MTJs leads to a loss of echo state property. 
To examine this consideration, the Lyapunov exponent is also evaluated, which clarifies the presence of echo state property. 
It is revealed that the boundary of the small-capacity region corresponds to the edge of echo state property. 
These results provide a procedure to achieve high computational ability of ASI, i.e., the edge of echo state property can be estimated from the saturation magnetization curve, and large memory capacities are obtained outside the edge. 


\begin{figure}
\centerline{\includegraphics[width=1.0\columnwidth]{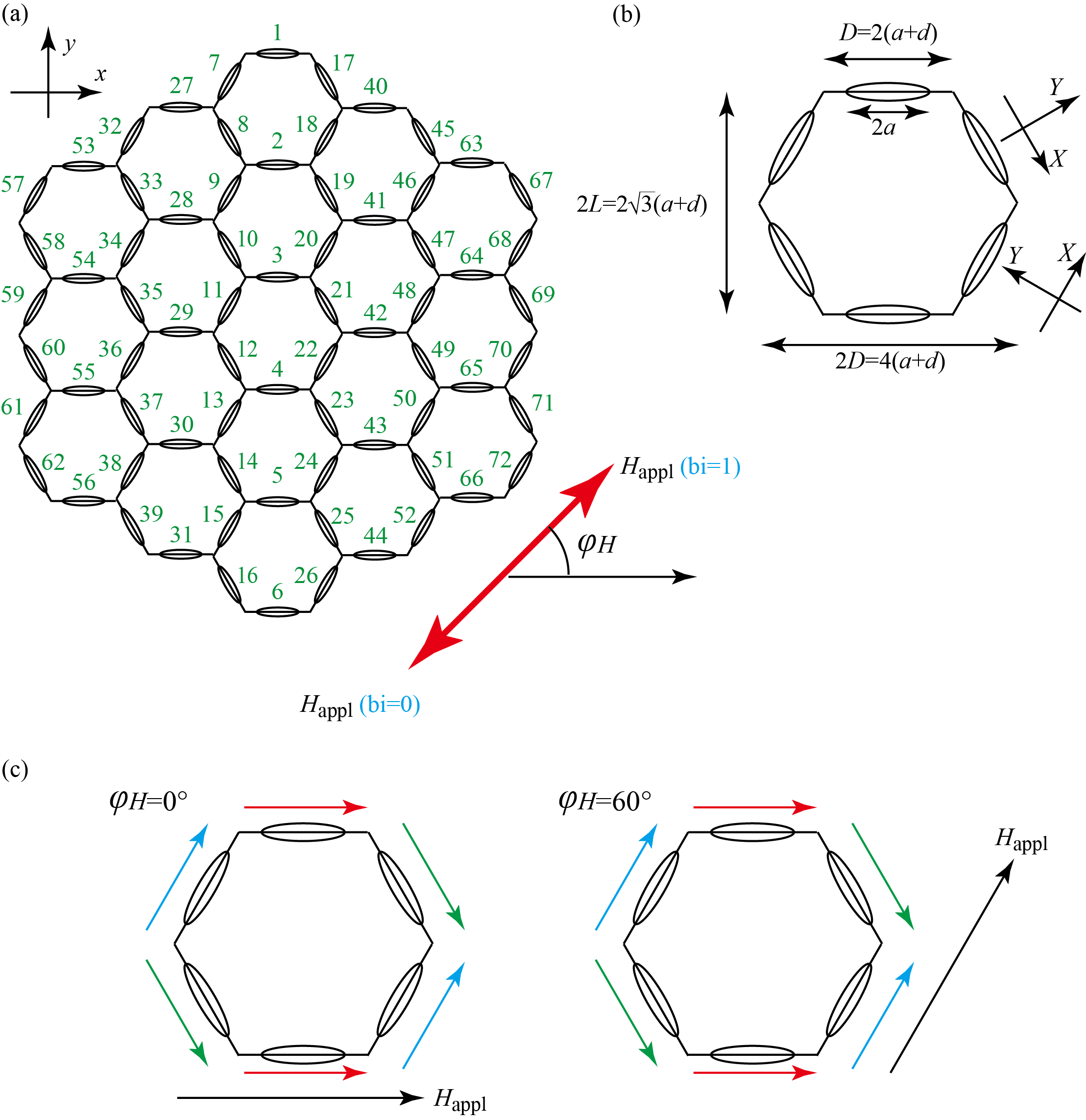}}
\caption{
            (a) Schematic illustration of ASI consisting of $72$ MTJ cells. 
                 Input signal is an external magnetic field with a strength $H_{\rm appl}$ and an angle $\varphi_{H}$ measured from the $x$ axis. 
                 The symbol ${\rm bi}$ represents the value of binary inputs used in the evaluation of memory capacities. 
            (b) Schematic illustration of one hexagon. 
                 The values of the long-axis radius and the distance from two MTJs to a cross point of their long axes is denoted as $d$. 
                 Schematic illustration of local coordinate $XYZ$ is also shown, where the $X$ axis is parallel to the long axis of an MTJ.
            (c) Magnetization alignments for the cases of $\varphi_{H}=0^{\circ}$ and $60^{\circ}$. 
         \vspace{-3ex}}
\label{fig:fig1}
\end{figure}


\section*{System description}


\subsection*{Description of artificial spin ice}

In Fig. \ref{fig:fig1}(a), we show a schematic view of ASI consisting of elliptical-shaped MTJs aligned in $xy$ plane \cite{kubota23}. 
The number of the MTJs is $N=72$, according to the experiment \cite{kubota24}.  
In this work, we use an external magnetic field with the strength $H_{\rm appl}$ which is applied in the direction having the angle $\varphi_{H}$ measured from the $x$ axis as input for a recognition task of time-dependent data; see also the following sections. 
Although the previous simulation assumes to utilize spin-transfer torque effect \cite{slonczewski96,berger96} for electrical manipulation of the magnetization states in an ASI, it is difficult to achieve because such a manipulation requires a thin ferromagnetic layer in an MTJ while the dipole interaction between MTJs become strong when a volume of an MTJ is large; this dilemma should be solved in future. 
The detection of the magnetization state via magnetoresistance effect, on the other hand, ca be experimentally achieved \cite{kubota24}; therefore, in principle, we can electrically detect the magnetization direction in each MTJ independently. 
Figure \ref{fig:fig1}(b) shows a schematic view of one hexagon in the ASI. 
The radii of the elliptical plane along the long and short axes are denoted as $a$ and $b$, while the distance from two MTJs to a cross point of their long axes is denoted as $d$. 
In the following calculation, we assume that the radii of the $i$th ($i=1,2,\cdots,N$) MTJ are randomly distributed around their designed values, $a$ and $b$, as $a_{i}=a\left(1+\sigma \xi_{ai}\right)$ and $b_{i}=b\left(1+\sigma\xi_{bi}\right)$, where $\xi_{ai}$ and $\xi_{bi}$ are uniformly distributed random values in the range of $-1 < \xi_{ai},\xi_{bi} < 1$ and the dimensionless parameter $\sigma$ determines the magnitude of the randomness of the MTJs' size. 
This is because a dispersion of the MTJs' size is unavoidable in experiments \cite{kubota23}. 
The thickness of the MTJs is assumed to be common. 
The values of the parameters are summarized in Methods for numerical method solving the LLG equation. 
We also introduce a local coordinate $XYZ_{i}$ for the latter discussion, where the $X_{i}$ and $Y_{i}$ axes are parallel to the $i$th MTJ. 
In the following, we call the $xyz$ and $XYZ$ coordinates as the global (or experimental) and local coordinates, respectively. 
In the experiment \cite{kubota24}, the magnetization directions of MTJs in the $x$ direction (in global coordinate) are measured. 
The local coordinate is, however, sometimes convenient to catch the dynamical behavior of the magnetization, define demagnetization coefficients, and so on. 
Therefore, we use both coordinates, depending on the situation. 

Since a honeycomb structure is unchanged even by rotating it around a perpendicular axis with $60^{\circ}$, one might consider that the present ASI also has the same symmetry. 
This is, however, not true due to two factors. 
First, we introduce the randomness in the MTJs size to reflect an experimental dispersion \cite{kubota23}, as mentioned above. 
Second, the presence of the magnetization vector breaks the structural symmetry. 
To clarify this point, we show the initial magnetization alignments of the magnetizations for the case of $\varphi_{H}=0^{\circ}$ and $60^{\circ}$; see Fig. \ref{fig:fig1}(c). 
Here, the magnetization directions of MTJs are indicated by arrows. 
We note that the magnetization directions are common for both cases. 
One can, however, notice that the magnetization alignment for $\varphi_{H}=0^{\circ}$ does not become identical to that for $\varphi_{H}=60^{\circ}$. 
This is because we use a common initial condition of the magnetization alignment in the $xy$ plane for two cases. 
Regarding these facts, the results shown below will be different for a certain $\varphi_{H}$ and $\varphi_{H}+60^{\circ}$, although the difference may be small. 


\subsection*{LLG equation}

The magnetization dynamics is evaluated by solving the LLG equation for the magnetization. 
We denote the unit vector pointing in the magnetization direction of the $i$th MTJ as $\mathbf{m}_{i}$. 
The LLG equation of $\mathbf{m}_{i}$ is given by 
\begin{equation}
  \frac{d \mathbf{m}_{i}}{dt}
  =
  -\gamma
  \mathbf{m}_{i}
  \times
  \left(
    \mathbf{H}_{\rm appl}
    +
    \mathbf{H}_{{\rm shape},i}
    +
    \mathbf{H}_{{\rm dip},i}
  \right)
  +
  \alpha 
  \mathbf{m}_{i}
  \times
  \frac{d\mathbf{m}_{i}}{dt},
  \label{eq:LLG}
\end{equation}
where the values of the gyromagnetic ratio $\gamma$ and the Gilbert damping constant $\alpha$ are assumed to be common for all MTJs and are given by $1.764\times 10^{7}$ rad/(Oe s) and $0.01$, respectively. 
The external magnetic field $\mathbf{H}_{\rm appl}$ is also commonly applied to the all MTJs. 
The shape magnetic anisotropy field is 
\begin{equation}
  \mathbf{H}_{{\rm shape},i}
  =
  -4\pi M 
  \sum_{U=X,Y,Z}
  N_{i,U}
  m_{i,U}
  \mathbf{e}_{i,U}, 
\end{equation}
where $\mathbf{e}_{i,U}$ ($U=X,Y,Z$) is the unit vector defining the local coordinate $XYZ_{i}$ mentioned above. 
Apparently, using the local coordinate is useful to express the shape demagnetization field. 
The demagnetization coefficient, $N_{i,U}$, along the $U$ direction is estimated numerically, according to the formula in Ref. \cite{beleggia05}, by using the radii, $a_{i}$ and $b_{i}$, and the thickness of the $i$th MTJ. 
The value of the saturation magnetization $M$ is $1500$ emu/c.c. 
Then, for an ideal MTJ ($a_{i}=a$ and $b_{i}=b$), $N_{iX}=0.040...$, $N_{Y}=0.154...$, and $N_{Z}=0.804...$, or equivalently, $4\pi MN_{X}\simeq 766$, $4\pi MN_{Y}\simeq 2919$, and $4\pi MN_{Z}\simeq 15165$ Oe. 
The dipole field $\mathbf{H}_{{\rm dip},i}$ is defined as 
\begin{equation}
  \mathbf{H}_{{\rm dip},i}
  =
  \sum_{j (\neq i) = 1}^{N}
  \mathbf{H}_{{\rm dip},ij}, 
\end{equation}
where $\mathbf{H}_{{\rm dip},ij}$ is the stray magnetic field from the $j$th MJT. 
We note that the magnitude of $\mathbf{H}_{{\rm dip},ij}$ is on the order of $100$ Oe at the most for the present parameters; see also Methods for the numerical method solving the LLG equation. 
The initial conditions of the magnetization are the alignments shown in Fig. \ref{fig:fig1}(c), where we assume that each magnetization is assumed to be parallel to the long axis of the MTJ. 


\begin{figure}
\centerline{\includegraphics[width=1.0\columnwidth]{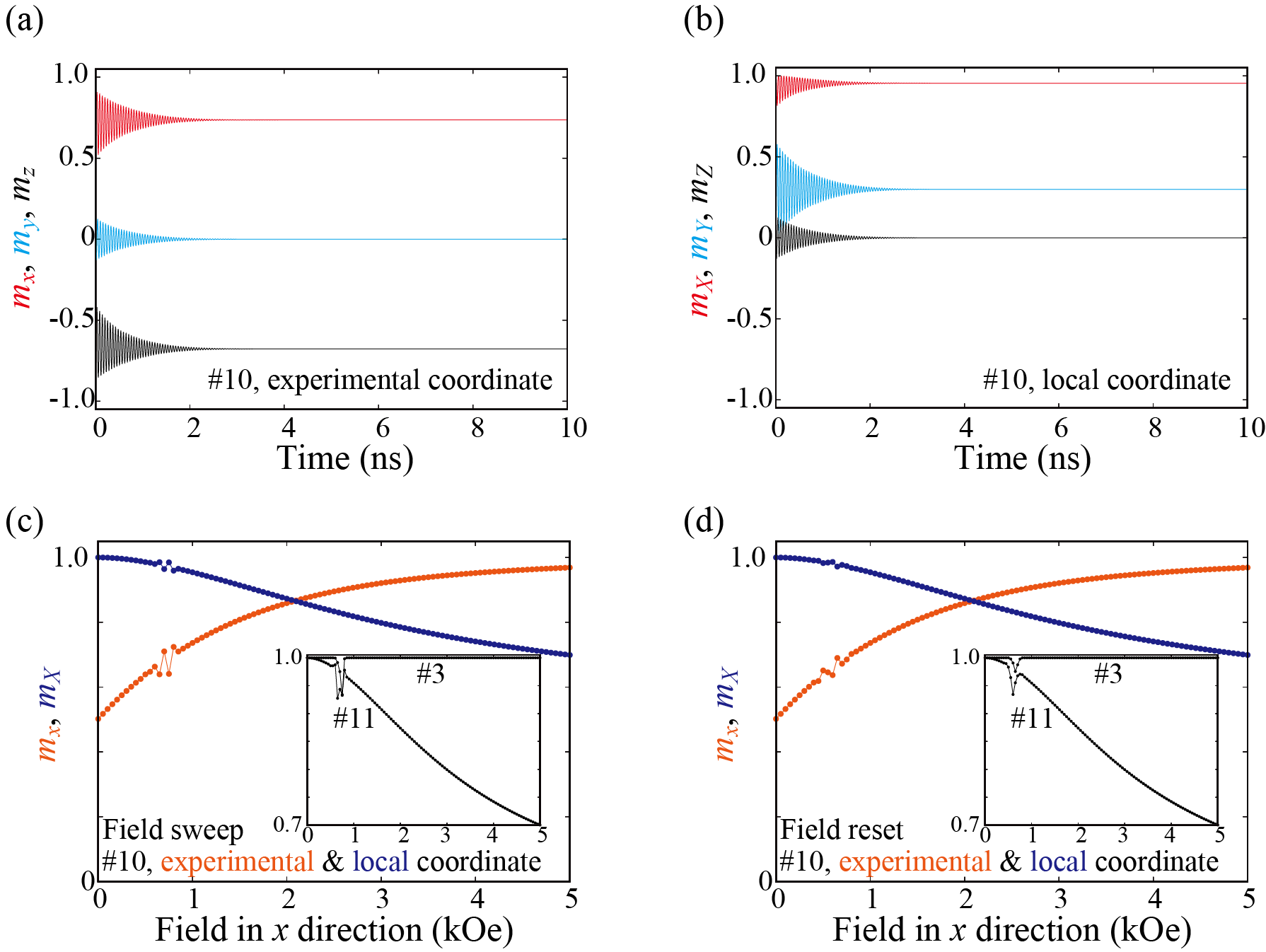}}
\caption{
            Time evolution of the magnetization in $10$th MTJ [see Fig. \ref{fig:fig1}(a)] in (a) global (experimental) and (b) local coordinate in the presence of an external magnetic field, $H_{\rm appl}=1.0$ kOe and $\varphi_{H}=0^{\circ}$.
            Saturated values of $m_{x}$ and $m_{X}$ with respect to various $H_{\rm appl}$ are summarized in (c) and (d). 
            In (c), the magnitude $H_{\rm appl}$ of the external magnetic field is swept while that in (d) is reset.
            The insets in (c) and (d) show the saturated values of $m_{X}$ in 3rd and 11th MTJs [see Fig. \ref{fig:fig1}(a)]. 
         \vspace{-3ex}}
\label{fig:fig2}
\end{figure}


Before moving to the physical reservoir computing, we investigate the role of the dipole interaction by solving Eq. (\ref{eq:LLG}) for various $H_{\rm appl}$. 
Figures \ref{fig:fig2}(a) and \ref{fig:fig2}(b) show time evolution of the magnetization in the $10$th MTJ [see Fig. \ref{fig:fig1}(a)] in (a) global (experimental) and (b) local coordinate. 
Recall that the magnetization is assumed to be parallel to the long-axis direction of the MTJ at the initial time; therefore, $m_{X}$, which is the $X$ component of the magnetization in the local coordinate, satisfies $|m_{X}|=1$ at the initial state. 
The magnitude $H_{\rm appl}$ and the angle $\varphi_{H}$ of the external magnetic field are $1.0$ kOe and $0^{\circ}$, respectively. 
As shown in the figure, the magnetization direction is eventually saturated after a few nanoseconds. 
We repeat similar calculations for various magnitude $H_{\rm appl}$ with a fixed angle $\varphi_{H}=0^{\circ}$. 
Figures \ref{fig:fig2}(c) and \ref{fig:fig2}(d) summarize the values of the saturated $m_{x}$ [$x$ component of the magnetization in the global (experimental) coordinate] and $m_{X}$ [$X$ component in the local coordinate]. 
In Fig. \ref{fig:fig2}(c), $H_{\rm appl}$ is swept from small to large value. 
In this case, we evaluate the saturation value of the magnetization under a certain value of $H_{\rm appl}$, and then, using this saturated value as a new initial condition, the saturated value of the magnetization under a slightly larger value of $H_{\rm appl}$ is evaluated. 
In Fig. \ref{fig:fig2}(d), on the other hand, the initial state of the magnetization is reset to the original state, i.e., parallel to the long-axis direction, for all $H_{\rm appl}$. 
Figure \ref{fig:fig2}(c) may be suitable for a comparison with experimental works, where, for example, a magnetization curve is often evaluated by using the sweep magnetic field \cite{kubota23}. 
Figure \ref{fig:fig2}(d), on the other hand, may be useful to catch the difference of the saturated value of the magnetization under the common initial condition. 
In both Figs. \ref{fig:fig2}(c) and \ref{fig:fig2}(d), $m_{x}$ becomes close to $+1$ because the external magnetic field points to the $x$ direction in this case ($\varphi_{H}=0^{\circ}$). 
Note that these figures indicate an appearance of non-monotonic behavior near $H_{\rm appl}\simeq 800$ Oe. 
We confirmed that such non-monotonic behavior also appears in the other MTJs; see the insets of Figs. \ref{fig:fig2}(c) and \ref{fig:fig2}(d), where the values of $m_{X}$ in the $3$rd and $11$th MTJs are shown [see Fig. \ref{fig:fig1}(a)]. 
Such a non-monotonic behavior does not appear when the dipole interaction is absent and thus, each MTJ independently reacts to the external magnetic field. 
Accordingly, the appearance of the non-monotonic behavior in the magnetization saturation to the $x$ direction originates from the dipole interaction between MTJs. 
It is, unfortunately, difficult to derive an analytical formula of the magnetic field corresponding to the non-monotonic saturation of the magnetization because of the complexity of the magnetic energy; see Methods for numerical method solving the LLG equation. 
It depends on both the shape magnetic anisotropy field and the dipole interaction. 
Although an analytical estimation is difficult, such a field strength will be estimated even in experiments by measuring saturation magnetization curve similar to Figs. \ref{fig:fig2}(c) and \ref{fig:fig2}(d).  
As we will see below, a drastic change of memory function in ASI appears around this field magnitude. 


\section*{Results}


\subsection*{Evaluation of memory capacities}

The short-term memory and parity check capacities, denoted as $C_{\rm STM}$ and $C_{\rm PC}$ in the following, respectively, are evaluated by applying random binary input ${\rm bi}_{\ell}=0$ or $1$ ($\ell=1,2,\cdots$) to the ASI as the external magnetic field as 
\begin{equation}
  \mathbf{H}_{\rm appl}
  =
  H_{\rm appl}
  \left(
    2 {\rm bi}_{\ell}
    -
    1
  \right)
  \left(
    \cos\varphi_{H}
    \mathbf{e}_{x}
    +
    \sin\varphi_{H}
    \mathbf{e}_{y}
  \right), 
  \label{eq:field}
\end{equation}
i.e., $\mathbf{H}_{\rm appl}=-H_{\rm appl}(\cos\varphi_{H}\mathbf{e}_{x}+\sin\varphi_{H}\mathbf{e}_{y})$ for ${\rm bi}_{\ell}=0$ and $\mathbf{H}_{\rm appl}=H_{\rm appl}(\cos\varphi_{H}\mathbf{e}_{x}+\sin\varphi_{H}\mathbf{e}_{y})$ for ${\rm bi}_{\ell}=1$ [see also Fig. \ref{fig:fig1}(a)]. 
The random input ${\rm bi}_{\ell}$ is constant during a pulse width $t_{\rm p}$. 
While the detail of the evaluation methods of the memory capacities is summarized in Methods, we briefly recall that the short-term memory and parity check capacities quantify the number of target data a reservoir can recognize. 
The target data for the evaluation of the short-term memory capacity is 
\begin{equation}
  z_{\ell,D}^{\rm STM}
  =
  {\rm bi}_{\ell-D},
  \label{eq:target_STM}
\end{equation}
i.e., the target data is the input signal itself. 
In other words, the short-term memory capacity characterize the number of the input data a resevoir can recognize as is. 
An integer $D(=0,1,2,\cdots)$ is called as delay, characterizing the order of the past input data. 
In this work, the short-term memory capacity is defined from memory function for $D=1,2,\cdots$; see also Methods for the evaluation method of the memory capacity. 
Note also that the memory capacities are the sum of component-wise capacities, which quantify the reproducibility of the target data with a given $D$; see the same Methods. 
The target data for the evaluation of the parity-check capacity is given by 
\begin{equation}
  z_{\ell,D}^{\rm PC}
  =
  \sum_{m=0}^{D}
  {\rm bi}_{\ell-D+m}\ \ \ \ ({\rm mod}\ 2).
  \label{eq:target_PC}
\end{equation}
The parity-check capacity is one type of nonlinear memory capacity. 
For simplicity, we call the short-term memory and parity-check capacities as memory capacities when it is unnecessary to distinguish them. 
The values of the parameters (for example, the number of the training data) for the evaluation of the short-term memory and parity check capacities are summarized in Methods. 
In the following, we show these memory capacities for various values of the parameters. 



\begin{figure}
\centerline{\includegraphics[width=1.0\columnwidth]{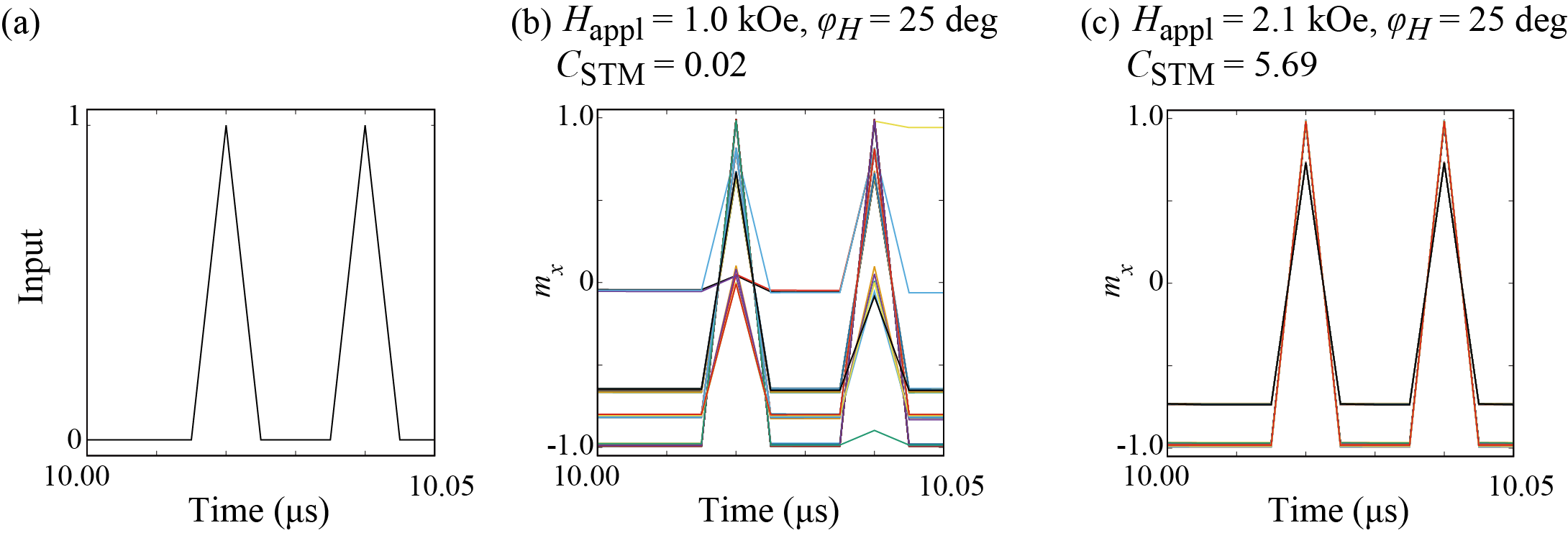}}
\caption{
            (a) An example of binary inputs and $m_{x}$ from MTJs for the conditions corresponding to (b) small $C_{\rm STM}(\simeq 0.02)$ and (c) large ($C_{\rm SMT}\simeq 5.69$) in Fig. \ref{fig:fig3}(a). 
            Explicitly, $H_{\rm appl}$ and $\varphi_{H}$ are (b) $H_{\rm appl}=1.0$ kOe and $\varphi_{H}=25^{\circ}$ and (c) $H_{\rm appl}=2.1$ kOe and $\varphi_{H}=25^{\circ}$. 
         \vspace{-3ex}}
\label{fig:fig3}
\end{figure}


Figure \ref{fig:fig3}(a) shows an example of the input binary data for the pulse width of $5$ ns, while Figs. \ref{fig:fig3}(b) and \ref{fig:fig3}(c) show the values of $N$-$m_{x}$ used for the evaluation of the memory capacities, where (b) $H_{\rm appl}=1.0$ kOe and $\varphi_{H}=25^{\circ}$ and (c) $H_{\rm appl}=2.1$ kOe and $\varphi_{H}=25^{\circ}$. 
As mentioned below, the short-term memory capacity corresponding to Fig. \ref{fig:fig3}(b) is small, while the condition of the magnetic field for Fig. \ref{fig:fig3}(c) gives the maximum value of the short-term memory capacity. 
In Fig. \ref{fig:fig3}(b), the magnetizations can show various states because the magnetic field strength is relatively small. 
Note that some of the magnetizations show non-identical behavior with respect to the same series of inputs. 
For example, Fig. \ref{fig:fig3}(a) shows two common series, where the random binary inputs change from $0$ to $1$ during $t=10.015$ $\mu$s and $t=10.020$ $\mu$s and during $t=10.025$ $\mu$s to $t=10.030$ $\mu$s. 
Similarly, the input changes from $1$ to $0$ during $t=10.020$ $\mu$s to $10.025$ $\mu$s and during $10.040$ $\mu$s and $t=10.045$ $\mu$s. 
The output data are, however, different for these series of the inputs data. 
For example, all the magnetization deviates from $m_{x}=-1.0$ when the input changes from $0$ to $1$ during $t=10.015$ $\mu$s and $t=10.020$ $\mu$s. 
Some magnetizations, however, remain near $m_{x}=-1.0$ when the input changes from $0$ to $1$ during $t=10.025$ $\mu$s and $t=10.030$ $\mu$s. 
Such a different response with respect to the same series of the input data will lead to a fail of learning and result in the component-wise capacity for small delay $D$. 
It may, on the other hand, contribute to the component-wise capacity for a large $D$; see also Methods for the evaluation of the memory capacities. 
We should note that the value of $H_{\rm appl}$ in Fig. \ref{fig:fig3}(b) is close to the magnetic field strength where the saturation curve of the magnetization becomes non-monotonic ($H_{\rm appl}\simeq 800$ Oe); see Figs. \ref{fig:fig2}(a) and \ref{fig:fig2}(b). 
It implies that the dipole interaction between MTJs lead to non-identical response of the magnetizations with respect to the same series of the input data. 
In contrast, in Fig. \ref{fig:fig3}(c), the magnetizations show almost the same response to the same series of the input data. 
Hence, this case is expected to recognize the target data for small delay $D$ and result in a relatively large memory capacities. 

\begin{figure}
\centerline{\includegraphics[width=1.0\columnwidth]{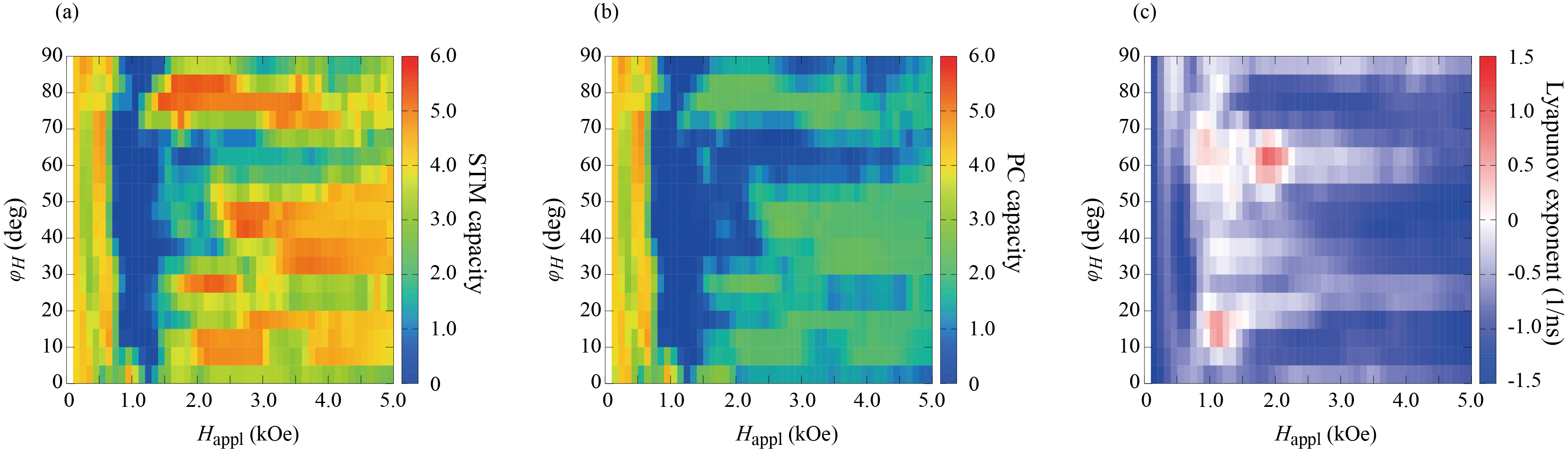}}
\caption{
         (a) Short-term memory capacity, (b) parity-check capacity, and (c) Lyapunov exponent for various magnetic field strength $H_{\rm appl}$ and angle $\varphi_{H}$.
         \vspace{-3ex}}
\label{fig:fig4}
\end{figure}

Figures \ref{fig:fig4}(a) and \ref{fig:fig4}(b) summarize the dependence of the short-term memory and parity-check capacities on the magnetic field strength $H_{\rm appl}$ and the angle $\varphi_{H}$. 
The values of the memory capacities are drastically changed, depending on the values of the parameters.  
For example, the short-term memory capacity is only $0.02$ for the condition in Fig. \ref{fig:fig3}(b) ($H_{\rm appl}=1.0$ kOe and $\varphi_{H}=25^{\circ}$), while it is maximized to $5.69$ for the condition in Fig. \ref{fig:fig3}(c) ($H_{\rm appl}=2.1$ kOe and $\varphi_{H}=25^{\circ}$). 
In particular, both capacities become small for the magnetic field strength near $H_{\rm appl}=800$ Oe. 
Recall again that this strength is close to the value at which non-monotonic saturation behavior is observed due to the dipole interaction; see Figs. \ref{fig:fig2}(c) and \ref{fig:fig2}(d). 
A large capacity appear near this small capacity condition, i.e., there is an edge between small and large capacities. 
In the next section, we investigate its origin by evaluating the Lyapunov exponent. 


\subsection*{Evaluation of  Lyapunov exponent and estimation of echo state property}

We evaluate the Lyapunov exponent $\varLambda$ to investigate the origin of the edge of small and large memory capacities in Figs. \ref{fig:fig4}(a) and \ref{fig:fig4}(b). 
The evaluation method of the Lyapunov exponent is summarized in Methods. 
Firstly, let us explain the motivation to evaluate the Lyapunov exponent. 


Physical reservoir computing works well when the output data from physical reservoir is solely determined by the input data and becomes independent of the initial state of the reservoir \cite{kubota21}. 
In other words, physical reservoir should show the same output with respect to the same input. 
This property is called echo state property. 
Whether physical reservoir can have the echo state property or not depends on the system parameters and the input characters (pulse strength and width, and so on). 
In some cases, the computational ability of physical reservoir is often maximized near the edge of echo state property, where the edge is a boundary of the system parameters and/or the input characters at which the echo state property is lost or recovered. 
The edge of echo state property is sometimes identical to the edge of chaos, which is a boundary between chaotic and ordered dynamics in physical reservoir. 
It should be, however, noted that these edges are not necessarily the same. 
In addition, while the edge of chaos has been often regarded as an ideal state for recurrent neural network \cite{bertschinger04}, it is recently mentioned by Jaeger in Foreword of Ref. \cite{nakajima21} that the edge of echo state property, not the edge of chaos, is a suitable state; it implies that the edge of chaos is a suitable state for the computation when it overlaps with the edge of echo state property. 
In addition, Jaeger points out that even this condition is limited, i.e., some tasks might be solved well at the edge of echo state property but the edge of echo state property is not always a best state for the computation. 
Based on this argument, the relationship between the short-term memory capacity, echo state property, and chaos for physical reservoir computing using spin-torque oscillators was recently investigated \cite{yamaguchi23}. 
We extend this previous work to ASI here and study the role of echo state property on the memory capacities. 
For this purpose, the Lyapunov exponent is evaluated due to the following reason. 


The Lyapunov exponent $\varLambda$ is an expansion rate of two dynamical responses with an infinitesimally different initial conditions. 
When the Lyapunov exponent is negative, two solutions saturate to a same state. 
When the Lyapunov exponent is zero, on the other hand, the difference between two solutions are kept to be constant. 
In the case of ferromagnetic system, for example, a magnetization relaxation (or switching) to a fixed stable state corresponds to the dynamics with $\varLambda<0$. 
On the other hand, an auto-oscillation of magnetization in a spin-torque oscillator corresponds to the dynamics with $\varLambda=0$ because, for example, if two oscillators have slightly different initial phases, the phase difference is kept to be constant due to the periodicity of the dynamics; in this case, the dynamical response depends on the initial phase. 
Recalling that the echo state property means that the dynamical response is independent of the initial state, the edge of echo state property can be defined as a boundary between negative and zero Lyapunov exponents. 
Therefore, the evaluation of the Lyapunov exponent will clarify the relationship between the echo state property and memory capacities. 


Figure \ref{fig:fig4}(c) show the Lyapunov exponent of the ASI for various $H_{\rm appl}$ and $\varphi_{H}$. 
Comparing it with Figs. \ref{fig:fig4}(a) and \ref{fig:fig4}(b), we notice that small memory capacities appear when the Lyapunov exponent is non-negative. 
Therefore, we conclude that the boundary between the small and large memory capacities is the edge of echo state property, which supports the argument by Jaeger \cite{nakajima21} mentioned above. 
This is also consistent with the result shown in Fig. \ref{fig:fig3}. 
When the output data shows different responses with respect to the same series ($01$ or $10$) input data in Fig. \ref{fig:fig3}(b), it means that the echo state property for $D=1$ is lost, and as a result, the short-term memory capacity is small. 
Summarizing these results, the dipole interaction between MTJs results in non-monotonic saturation of the magnetizations in ASI at a certain strength of the external magnetic field, and the memory capacities becomes small near this field strength due to the loss of echo state property. 


At the end of this section, we give some comments on the positive Lyapunov exponent in Fig. \ref{fig:fig4}(c). 
An existence of a positive Lyapunov exponent is sometimes regarded as the presence of chaotic dynamics \cite{alligood97,strogatz01,ott02}. 
We, however, do not consider that the positive Lyapunov exponent in Fig. \ref{fig:fig4}(c) indicate the presence of chaotic behavior in ASI. 
This is because the magnetizations in MTJs tend to saturate to certain directions by the application of the external magnetic field, as shown in Figs. \ref{fig:fig2}(a) and \ref{fig:fig2}(b), while chaos is a sustainable dynamics and its dynamical trajectory does not saturate. 
The presence of the positive Lyapunov exponent arises from the fact that there are several magnetization alignments corresponding to local minima of magnetic energy, and which minimum the system settles on depends on various factors including the initial state of the magnetizations. 
This point may be differently explained in terms of attractor in nonlinear science \cite{alligood97,strogatz01,ott02}. 
An attractor is a region in a phase space, toward which a system evolves; thus, if the dynamical state initially locates inside an attractor, the system remains in it. 
A local minimum of energy landscape is a typical attractor, called a point attractor, because a system will fall into the minimum due to energy dissipation (relaxation). 
There are also various kinds of attractors, such as periodic and chaotic (strange) attractors \cite{alligood97,strogatz01,ott02}. 
In the present system, the magnetization dynamics shown in Figs. \ref{fig:fig2} and \ref{fig:fig3} indicate that there are several point attractors corresponding to the local minima of magnetic energy, and the ASI saturates to one of them. 
A positive Lyapunov exponent appears when two systems with slightly different initial conditions fall into different point attractors; in this case, although the distance between two systems is expanded, the dynamics cannot be regarded as chaos. 
Therefore, we do not consider that chaotic dynamics exists in this work; accordingly, edge of chaos is also absent in this work. 
In addition, the fact that the region corresponding to the positive Lyapunov exponent in Fig. \ref{fig:fig4}(c) is small and does not fit with the boundary between the small and large memory capacities again confirms that the edge of echo state property, not the edge of chaos, determines the boundary. 



\section*{Conclusion}

In summary, magnetization dynamics in an ASI and its memory capacities for physical reservoir computing were investigated by numerical simulation of the LLG equation. 
A non-monotonic saturation of the magnetization with respect to an application of an external magnetic field was observed, which originated from the dipole interaction between MTJs. 
Both the short-term memory and parity-check capacities become quite small for the magnetic field strength around which such a non-monotonic behavior appears. 
By evaluating the Lyapunov exponent, as well as investigating the temporal magnetization dynamics, it was found that the small memory capacities were due to the loss of echo state property. 
In other words, the boundary between the small and large memory capacities corresponds to the edge of echo state property. 





\section*{Methods}


\subsection*{Numerical method solving the LLG equation}

We apply the fourth-order Runge-Kutta method to the LLG equation with a time increment of $\Delta t=1.0$ ps. 
In this work, we use $a=200$ nm, $b=75$ nm and $d=20$ nm, while the thickness of the MTJs is assumed to be common for all the specimen and is $20$ nm \cite{kubota23}. 
The Mersenne Twister \cite{matsumoto98} for Fortran was used to generate random numbers, such as $\xi_{ai}$, $\xi_{bi}$, and ${\rm bi}_{\ell}$, in this work. 
For example, $\xi_{a1}$ ad $\xi_{b1}$ are $-0.326...$ and $-0.569...$. 
Using these random numbers, both the shape magnetic and dipole fields of MTJs become slightly random around their designed values, where the designed value means that all MTJs have the common radii, $a$ and $b$. 
Recall that the dispersion of the MTJs size is characterized by the dimensionless parameter $\sigma$. 
Assuming that the central positions of MTJs are located as in the case of an ideal honeycomb structure,  $\sigma$ should satisfy $\sigma \le d/a$ to avoid an overlap of MTJs. 
In this work, we use $\sigma=0.02$, i.e., the maximum difference of $a_{i}$ and $b_{i}$ from the designed value is $2$\%. 

Recall that the dipole field acting on the $i$th MTJ is the sum of the stray magnetic field from the other MTJs. 
The stray magnetic field $\mathbf{H}_{{\rm dip},ij}$ from the $j$th MTJ is numerically estimated by the method developed in Ref \cite{taniguchi23}, where the value of the stray magnetic field at the center of the $i$th MTJ is used as $\mathbf{H}_{{\rm dip},ij}$. 
In general, the stray magnetic field can be expressed as 
\begin{equation}
  \mathbf{H}_{{\rm dip},ij}
  =
  M
  \begin{pmatrix}
    I_{i,j,1,1} & I_{i,j,1,2} & I_{i,j,1,3} \\
    I_{i,j,2,1} & I_{i,j,2,2} & I_{i,j,2,3} \\
    I_{i,j,3,1} & I_{i,j,3,2} & I_{i,j,3,3}
  \end{pmatrix}
  \begin{pmatrix}
    m_{j,x} \\
    m_{j,y} \\
    m_{j,z}
  \end{pmatrix}, 
  \label{eq:H_dip_ij_example}
\end{equation}
where we use the global coordinate, for convenience. 
The values of the matrix elements, $I_{i,j,p,q}$ ($p,q,=1,2,3$) depend on the size ($a_{j}$ and $b_{j}$) of the $j$th MTJ and the relative position between the $i$th and $j$th MTJs \cite{taniguchi23}. 
For example, using $\xi_{a1}$ and $\xi_{b1}$ mentioned above and $M=1500$ emu/c.c., the values of the stray magnetic field from the $1$st ($i=1$) to the $7$th ($j=7$) MTJ are estimated to be $MI_{7,1,1,1}\simeq 51$, $MI_{7,1,1,2}\simeq 37$, $MI_{7,1,2,1}\simeq 81$, $MI_{7,1,2,2}\simeq 7$, and $MI_{7,1,3,3}\simeq -34$ Oe (the other matrix elements are zero). 
Recall that these MTJs are an example of the nearest MTJs [see Fig. \ref{fig:fig1}(a)]. 
Therefore, we mentioned in the main text that the stray magnetic field $\mathbf{H}_{{\rm dip},ij}$ from one MTJ is on the order of $100$ Oe at maximum. 
Note that an MTJ has four nearest MTJs (except MTJs around the edge of ASI), and thus, the strength of the dipole field $\mathbf{H}_{{\rm dip},i}$ can be about four times larger than that of $\mathbf{H}_{{\rm dip},ij}$ at maximum. 
We notice that the non-monotonic behavior of the saturation curves observed in Figs. \ref{fig:fig2}(c) and \ref{fig:fig2}(d) relates to the randomness of the MTJ size. 
In the absence of the randomness ($\sigma=0$), the non-monotonic behavior disappears. 
We consider that it comes from a cancellation of the stray magnetic field. 
For example, when we focus on the $3$rd MTJ in Fig. \ref{fig:fig1}(a), only the $x$ component of the stray magnetic fields generated by the surrounding MTJs remain finite, while the $y$ component becomes totally zero due to the cancellation, when the randomness is absent. 
The remaining $x$ component of the total stray magnetic field just enhances the applied magnetic field and does not lead any non-monotonic behavior in the saturation curve. 
When the randomness is finite, on the other hand, the total $y$ component becomes finite and gives the non-monotonic behavior.

The magnetic energy density is the sum of the Zeeman energy, shape magnetic anisotropy energy, and the dipole interactions, which are modeled as 
\begin{equation}
  E
  = 
  -\sum_{i=1}^{N}
  M \mathbf{m}_{i}
  \cdot
  \mathbf{H}_{\rm appl}
  +
  2\pi M^{2}
  \sum_{i=1}^{N}
  \sum_{U=X,Y,Z}
  N_{i,U}
  m_{i,U}^{2}
  -
  \sum_{i=1}^{N}
  M \mathbf{m}_{i}
  \cdot
  \mathbf{H}_{{\rm dip},i}. 
\end{equation}
Because of several reasons, such as randomness of the shape magnetic anisotropy and dipole fields and finite size of ASI, it is difficult to estimate the magnetic field strength around which the saturation magnetization curve shows non-monotonic behavior shown in Figs. \ref{fig:fig2}(c) and \ref{fig:fig2}(d). 


\subsection*{Evaluation method of memory capacities}

The evaluation method of the short-term memory and parity-check capacities consists of two steps, where we firstly evaluate weight by learning (training) and secondly evaluate the capacities by using the weight \cite{jaeger02,bertschinger04,dambre12,fujii17,tsunegi18,kubota21}. 
In the following, we summarize these procedures. 

The first procedure is learning (training). 
A series of pulse input signal $r_{\ell}$ ($\ell=1,2,\cdots,N_{\rm L}$) is injected into physical reservoir, where $N_{\rm L}$ is the number of the input signal for the determination of the weight. 
The suffix $\ell$ distinguishes the order of the input signal. 
A binary input signal $b_{\ell}$ is used in the present work, as in the case of Ref. \cite{tsunegi18}. 
Another choice of the input signal is a uniformly distributed random signal ($0 \le r_{\ell} \le 1$ or $-1 \le r_{\ell} \le 1$) \cite{kubota21}. 
The target data $z_{\ell,D}$ can be defined from the input signal $r_{\ell}$. 
The delay $D(=0,1,2,\cdots)$ quantifies the number of the past data used to define target data. 
Since physical reservoir computing aims to recognize past input signal from the present output signal, it is necessary to introduce the delay $D$ to distinguish the order of the past input signal. 
The target data of the short-term memory and parity-check capacities are shown in Eqs. (\ref{eq:target_STM}) and (\ref{eq:target_PC}). 
After defining the target data, the weight $w_{D,i}$ is estimated to minimize, 
\begin{equation}
  \sum_{\ell=1}^{N_{\rm L}}
  \left(
    \sum_{i=1}^{N+1}
    u_{\ell,i}
    w_{D,i}
    -
    z_{\ell,D}
  \right)^{2}
\end{equation}
where $u_{\ell,i}$ is the output data from the $i$th node (MTJ in the present work) with respect to the $\ell$th input signal. 
In this work, $m_{i,x}$ at time just before switching an input pulse input is used as $u_{\ell,i}$, i.e., when $\ell$th input is injected during time interval $t_{\ell} < t \le t_{\ell}+t_{\rm p}$, $m_{i,x}$ at $t=t_{\ell}+t_{\rm p}$ is used as $u_{\ell,i}$. 
The $N+1$th output data, $u_{\ell,N+1}$, is the bias term, $u_{\ell,N+1}=1$. 
The weight is obtained as  Moore-Penrose inverse matrix of $u_{\ell,i}$. 
Note that the value of the weight $w_{\rm D,i}$ depends on the target data; therefore, strictly speaking, it is necessary to add some index to $w_{\rm D,i}$ to distinguish the weights for the evaluation of the short-term memory and parity-check capacities. 
For simplicity, however, we omit such an index because the following procedure is commonly adopted for their evaluations. 
Recall that the target data $z_{\ell,D}$ for the short-term memory task is the random binary input ${\rm bi}_{\ell-D}$ and its example is shown in Fig. \ref{fig:fig3}(a). 
Also, the output data $u_{\ell,i}$ is $m_{i,x}$, as mentioned above, and their examples for $72$ MTJs are shown in Figs. \ref{fig:fig3}(b) and \ref{fig:fig3}(c).


\begin{figure}
\centerline{\includegraphics[width=1.0\columnwidth]{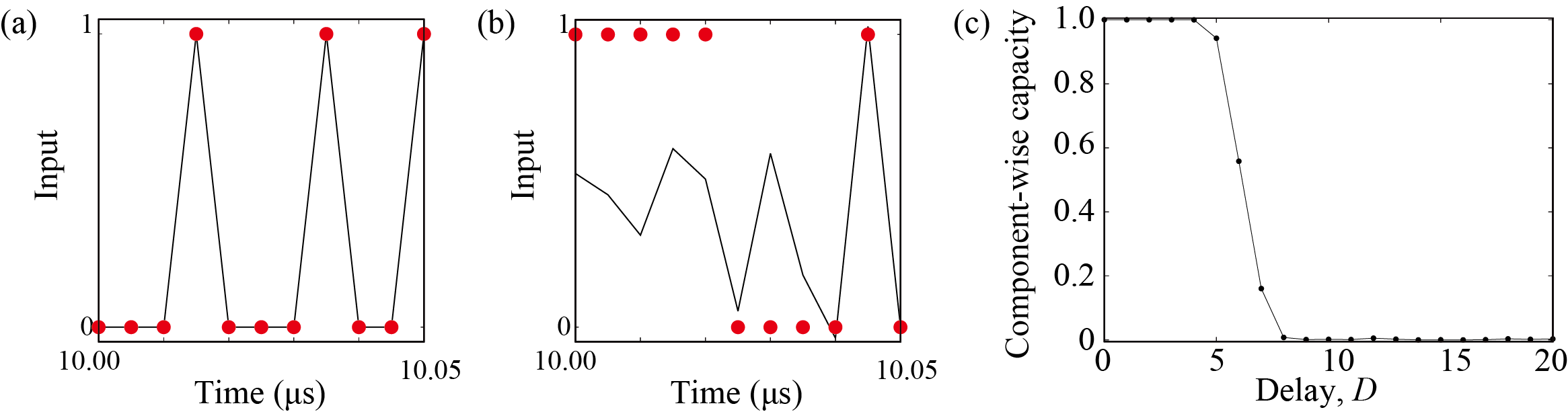}}
\caption{
         Examples of target data and system output for the evaluation of the short-term memory capacity with the delay of (a) $D=0$ and (b) $D=6$. 
         The values of the parameters are identical to those used in Fig. \ref{fig:fig3}(c). 
         (c) Component-wise capacity for the evaluation of the short-term memory capacity. 
         \vspace{-3ex}}
\label{fig:fig5}
\end{figure}


Next, we evaluate the memory capacities by injecting a different series of input signal $r_{n}^{\prime}$ ($n=1,2,\cdots,N_{\rm E}$), where the prime symbol is added to quantities to distinguish them from those used in learning. 
The number of the input signal, $N_{\rm E}$, is not necessarily the same as the number used in learning ($N_{\rm L}$). 
From the output data $u_{n,i}^{\prime}$ as the response to $r_{n}^{\prime}$ and using the weight determined by learning, system output is defined as 
\begin{equation}
  y_{n,D}^{\prime}
  =
  \sum_{i=1}^{N+1}
  u_{n,i}^{\prime}
  w_{D,i}. 
\end{equation}
If the learning is done well, the system output will reproduce the target data $z_{n,D}^{\prime}$ defined from $r_{n}^{\prime}$. 
In Fig. \ref{fig:fig5}, we show examples of the target data (red dots) and the system output (black line) for the evaluation of the short-term memory capacity ($z_{\ell,D}^{\rm STM}={\rm bi}_{\ell-D}$ and $u_{n,i}^{\prime}=m_{i,x}$) with the delays of (a) $D=0$ and (b) $D=6$. 
As shown, the target data with the small delay ($D=0$) is easily reproduced because the output data includes the past information within short time. 
For the large delay ($D=6$), however, the reproducibility becomes low because the relaxation dynamics of the magnetization erases the past information sufficiently before the present time. 
The reproducibility is quantified by the correlation coefficient, 
\begin{equation}
  {\rm Cor}(D)
  =
  \frac{\sum_{n=1}^{N_{\rm E}} \left( z_{n,D}^{\prime} - \langle z_{n,D}^{\prime} \rangle \right) \left( y_{n,D}^{\prime} - \langle y_{n,D}^{\prime} \rangle \right)}
    {\sqrt{ \sum_{n=1}^{N_{\rm E}} \left( z_{n,D}^{\prime} - \langle z_{n,D}^{\prime} \rangle \right)^{2} \sum_{n=1}^{N_{\rm E}} \left( y_{n,D}^{\prime} - \langle y_{n,D}^{\prime} \rangle \right)^{2} }}, 
    \label{eq:correlation}
\end{equation}
where the symbol $\langle \cdots \rangle$ means an average. 
The component-wise capacity for the target data $z_{n,D}^{\prime}$ is defined as 
\begin{equation}
  C(z_{n,D}^{\prime})
  =
  \left[
    {\rm Cor}(D)
  \right]^{2}.
\end{equation}
Figure \ref{fig:fig5}(c) shows the examples of the component-wise capacities for $D=0, 1,2,\cdots,20$, where the examples of the target data and the system output in Eq. (\ref{eq:correlation}) are shown in Figs. \ref{fig:fig5}(a) and \ref{fig:fig5}(b), as mentioned above. 
The component-wise capacity is unity when the system output completely reproduce the target data, while it tends to approach zero when the system output is largely different from the target data. 
The component-wise capacity is defined for each target data, $z_{n,D}^{\prime}$ ($D=0,1,2\cdots$). 
The memory capacity is defined as 
\begin{equation}
  C
  =
  \sum_{D=1}^{D_{\rm max}}
  C(z_{n,D}^{\prime}), 
\end{equation}
where $D_{\rm max}$ is the maximum delay. 
Note that we use the definition of the memory capacity in Ref. \cite{tsunegi18}, where the component-wise capacities from $D=1$ are used for the evaluation of the memory capacity. 
In some papers, however, the component-wise capacity for $D=0$ is included in the definition of the memory capacity \cite{fujii17}. 
The component-wise capacity often becomes small for a large delay $D$, except some cases \cite{yamaguchi20}, and thus, the value of the memory capacity is saturated when $D_{\rm max}$ is set to sufficiently large value. 
According to the definition of the component-wise capacity, for example, it is small in the case in Fig. \ref{fig:fig3}(b), where, as mentioned in the main text, some magnetizations show different response with respect to the same series of the input data, $01$ or $10$; in this case, the component-wise capacity even for $D=1$ becomes small. 
In fact, the short-term memory capacity of this case is sufficiently smaller than $1$, indicating that the ASI cannot recognize the input data injected even only one step before. 
At the same time, however, we should note that a different response with respect to two digits inputs may contribute to the component-wise capacity for a large $D(\ge 2)$ because such a different response may reflect a sufficiently past input signal. 


\begin{figure}
\centerline{\includegraphics[width=1.0\columnwidth]{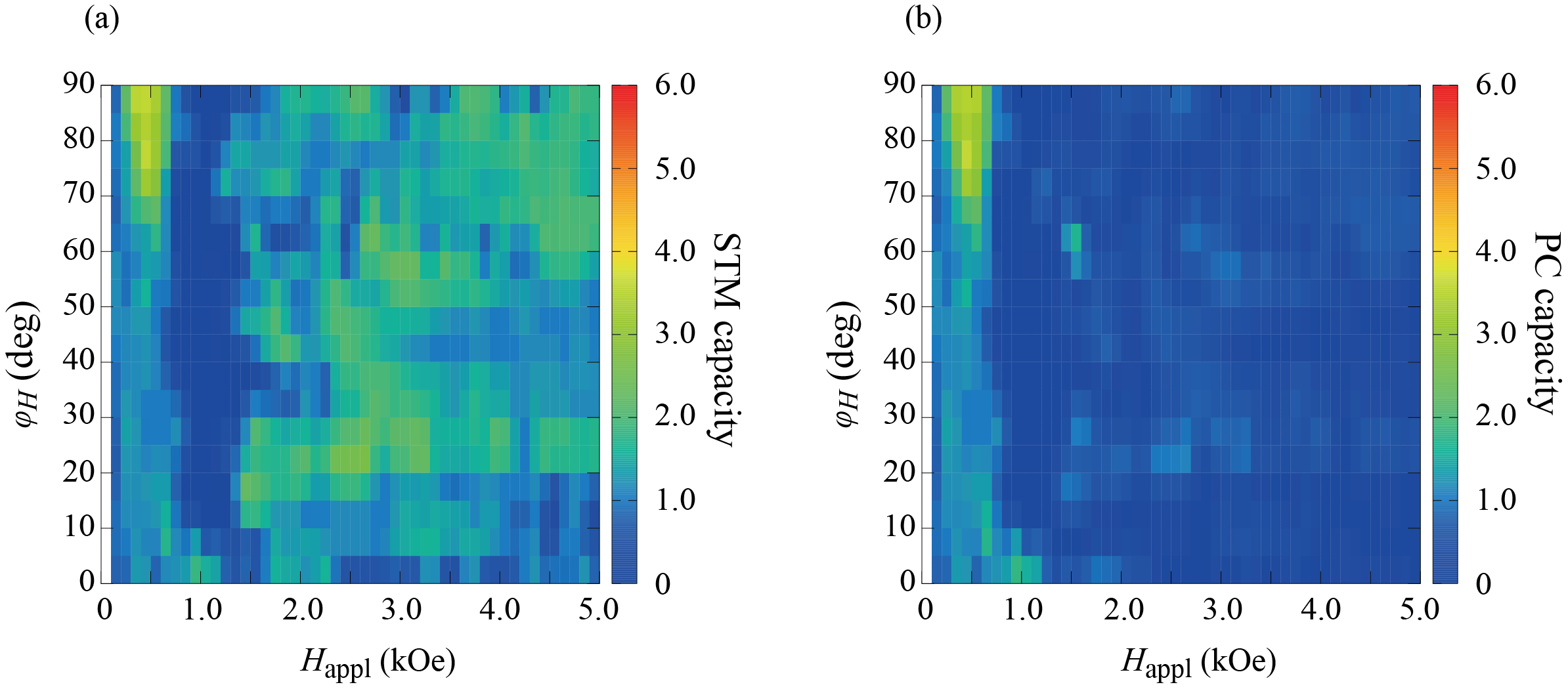}}
\caption{
         (a) Short-term memory and (b) parity-check capacities for various magnetic field strength $H_{\rm appl}$ and angle $\varphi_{H}$, where the precision of the system output $m_{x}$ is reduced by the method described in the text. 
         \vspace{-3ex}}
\label{fig:fig6}
\end{figure}


In the present work, we first solve the LLG equation with a fixed magnetic field from $t=0$ to $t=100$ ns, where the magnetic field is given by Eq. (\ref{eq:field}) and ${\rm bi}$ is fixed to $+1$. 
This procedure is to saturate the ASI to a stable state under the constant magnetic field. 
Then, $300$ random binary inputs are injected for washout, $N_{\rm L}=1000$ random binary inputs are injected  for learning, $300$ random binary inputs are injected for washout again, and $N_{\rm E}=1000$ random binary inputs are injected for the evaluation of the memory capacities. 


In the main text, the memory capacities are evaluated by solving the LLG equation with $64$bits Fortran code. 
Therefore, the output data are obtained with high precision. 
In such a case, for example, although the magnetization is almost saturated at $t=5$ ns in Fig. \ref{fig:fig2}(a), a small change of $m_{x}$ may possibly occur during the relaxation process (recall that the pulse width of the input signal is $t_{\rm p}=5$ ns). 
The learning and evaluation processes of the memory capacities might be affected by such a small change of $m_{x}$. 
A precision of experimental components \cite{kubota24} is, however, often small compared with that in the computational evaluation. 
Therefore, an attempt is made to evaluate the short-term memory and parity-check capacities by using the same output data used in Fig. \ref{fig:fig4} except for the fact that the value of the output data, $m_{i,x}$ is reduced to three digits; for example, if $m_{ix}$ is $m_{i,x}=0.12345\cdots$, $m_{ix}=0.12300...$ is used as the output. 
Figures \ref{fig:fig6}(a) and \ref{fig:fig6}(b) summarize the short-term memory and parity-check capacities evaluated by this reduced precision. 
The color scale is the same with that used in Figs. \ref{fig:fig4}(a) and \ref{fig:fig4}(b). 
We confirm that, although the values of the memory capacities become small, several characteristics, such as a change of the memory capacities near the magnetic field strength of $800$ Oe, show common tendencies for  Figs. \ref{fig:fig4} and \ref{fig:fig6}. 
Given the results so far, we believe that the results obtained by the numerical simulation in this work will be tested by future experiments. 



\subsection*{Evaluation method of Lyapunov exponent}

The Lyapunov exponent is evaluated by Shimada-Nagashima method \cite{shimada79}. 
In this method, the maximum Lyapunov exponent is obtained as an average of temporal Lyapunov exponent, which can be estimated from an instantaneous expansion rate of two infinitesimally separated magnetizations to the most expanded direction. 
We introduce the zenith and azimuth angles of the $k$th MTJ ($k=1,2,\cdots,N$), $\theta_{k}(t)$ and $\varphi_{k}(t)$, as $\mathbf{m}_{k}=(\sin\theta_{k}\cos\varphi_{k},\sin\theta_{k}\sin\varphi_{k},\cos\theta_{k})$ and evaluate their expansion rate \cite{taniguchi20,taniguchi24srep,taniguchi24}. 

Let us denote the initial time to evaluate the Lyapunov exponent as $t_{\rm init}$. 
We also denote the $k$th magnetization at $t=t_{\rm init}$ as $\mathbf{m}_{k}(t_{\rm init})$. 
At $t=t_{\rm init}$, we introduce $\mathbf{m}_{k}^{(1)}(t_{\rm init})=[\sin\theta_{k}^{(1)}\cos\varphi_{k}^{(1)},\sin\theta_{k}^{(1)}\sin\varphi_{k}^{(1)},\cos\theta_{k}^{(1)}]$, where the distance between $\mathbf{m}_{k}(t_{\rm init})$ and $\mathbf{m}_{k}^{(1)}(t_{\rm init})$ is infinitesimally small. 
For simplicity, we denote this distance as 
\begin{equation}
  \mathscr{D}_{k}^{(1)}[\mathbf{m}_{k}(t_{\rm init}),\mathbf{m}_{k}^{(1)}(t_{\rm init})]
  =
  \sqrt{
    \left[
      \theta_{k}(t_{\rm init})
      -
      \theta_{k}^{(1)}(t_{\rm init})
    \right]^{2}
    +
    \left[
      \varphi_{k}(t_{\rm init})
      -
      \varphi_{k}^{(1)}(t_{\rm init})
    \right]^{2}
  }. 
\end{equation}
Since $\mathbf{m}_{k}^{(1)}$ is introduced over all the MTJs, i.e., $k=1,2,\cdots,N$, we define the total distance as 
\begin{equation}
  \mathscr{D}^{(1)}(t_{\rm init})
  =
  \sum_{k=1}^{N}
  \mathscr{D}_{k}^{(1)}[\mathbf{m}_{k}(t_{\rm init}),\mathbf{m}_{k}^{(1)}(t_{\rm init})]. 
\end{equation}
The value of $\mathscr{D}^{(1)}(t_{\rm init})$ is fixed to be a small value, which is denoted as $\varepsilon$ and is $1.0\times 10^{-5}$ in this work. 
The values of $\theta_{k}^{(1)}(t_{\rm init})$ and $\varphi_{k}^{(1)}(t_{\rm init})$ are arbitrary by keeping the condition $\mathscr{D}^{(1)}(t_{\rm init})=\varepsilon$. 

The time evolution of $\mathbf{m}_{k}(t_{\rm init})$ and $\mathbf{m}_{k}^{(1)}(t_{\rm init})$ to $\mathbf{m}_{k}(t_{\rm init}+\Delta t)$ and $\mathbf{m}_{k}^{(1)}(t_{\rm init}+\Delta t)$ is obtained by solving the LLG equation. 
Note that their distance, 
\begin{equation}
  \mathscr{D}_{k}^{(1)}[\mathbf{m}_{k}(t_{\rm init}+\Delta t),\mathbf{m}_{k}^{(1)}(t_{\rm init}+\Delta t)]
  =
  \sqrt{
    \left[
      \theta_{k}(t_{\rm init}+\Delta t)
      -
      \theta_{k}^{(1)}(t_{\rm init}+\Delta t)
    \right]^{2}
    +
    \left[
      \varphi_{k}(t_{\rm init}+\Delta t)
      -
      \varphi_{k}^{(1)}(t_{\rm init}+\Delta t)
    \right]^{2}
  }, 
\end{equation}
is generally different from $\mathscr{D}_{k}^{(1)}[\mathbf{m}_{k}(t_{\rm init}),\mathbf{m}_{k}^{(1)}(t_{\rm init})]$. 
Therefore, introducing 
\begin{equation}
  \mathscr{D}^{(1)}(t_{\rm init}+\Delta t)
  =
  \sum_{k=1}^{N}
  \mathscr{D}_{k}^{(1)}[\mathbf{m}_{k}(t_{\rm init}+\Delta t),\mathbf{m}_{k}^{(1)}(t_{\rm init}+\Delta t)], 
\end{equation}
the expansion rate of the magnetizations from $t=t_{\rm init}$ to $t=t_{\rm init}+\Delta t$ is given by $\mathscr{D}^{(1)}(t_{\rm init}+\Delta t)/\varepsilon$. 
The temporal Lyapunov exponent at $t=t_{\rm init}+\Delta t$ is then defined as 
\begin{equation}
  \varLambda^{(1)}
  =
  \frac{1}{\Delta t}
  \ln
  \frac{\mathscr{D}^{(1)}(t_{\rm init}+\Delta t)}{\varepsilon}. 
\end{equation}

While the initial perturbations, $\theta_{k}^{(1)}(t_{\rm init})-\theta_{k}(t_{\rm init})$ and $\varphi_{k}^{(1)}(t_{\rm init})-\varphi_{k}(t_{\rm init})$, are arbitrary, the perturbation from $t=t_{\rm init}+\Delta t$ should be chosen so that $\mathbf{m}_{k}^{(n)}[t_{\rm init}+(n-1)\Delta t]-\mathbf{m}[t_{\rm init}+(n-1)\Delta t]$ ($n=2,3,\cdots$) points to the most expanded direction. 
This is a key idea of the Shimada-Nagashima method \cite{shimada79}, i.e., even if the initial perturbation is arbitrary, the distance between two solutions of equation of motion will naturally move to the most expanded direction. 
In the present case, at $t=t_{\rm init}+(n-1)\Delta t$ ($n=2,3,\cdots$), we define $\mathbf{m}_{k}^{(n)}[t_{\rm init}+(n-1)\Delta t]$ by introducing the zenith and azimuth angles as 
\begin{equation}
  \theta_{k}^{(n)}[t_{\rm init}+(n-1)\Delta t]
  =
  \theta_{k}[t_{\rm init}+(n-1)\Delta t]
  +
  \varepsilon
  \frac{\theta_{k}^{(n-1)}[t_{\rm init}+(n-1)\Delta t]-\theta_{k}[t_{\rm init}+(n-1)\Delta t]}{\mathscr{D}^{(n-1)}[t_{\rm init}+(n-1)\Delta t]},
\end{equation}
\begin{equation}
  \varphi_{k}^{(n)}[t_{\rm init}+(n-1)\Delta t]
  =
  \varphi_{k}[t_{\rm init}+(n-1)\Delta t]
  +
  \varepsilon
  \frac{\varphi_{k}^{(n-1)}[t_{\rm init}+(n-1)\Delta t]-\varphi_{k}[t_{\rm init}+(n-1)\Delta t]}{\mathscr{D}^{(n-1)}[t_{\rm init}+(n-1)\Delta t]},
\end{equation}
where $\mathscr{D}^{(n-1)}[t_{\rm init}+(n-1)\Delta t]$ is defined as 
\begin{equation}
\begin{split}
  \mathscr{D}^{(n-1)}[t_{\rm init}+(n-1)\Delta t]
  &=
  \sum_{k=1}^{N}
  \mathscr{D}_{k}^{(n-1)}\{\mathbf{m}_{k}[t_{\rm init}+(n-1)\Delta t],\mathbf{m}_{k}^{(n-1)}[t_{\rm init}+(n-1)\Delta t]\},
\\
  &=
  \sum_{k=1}^{N}
  \sqrt{
    \left\{
      \theta_{k}[t_{\rm init}+(n-1)\Delta t]
      -
      \theta_{k}^{(n-1)}[t_{\rm init}+(n-1)\Delta t]
    \right\}^{2}
    +
    \left\{
      \varphi_{k}[t_{\rm init}+(n-1)\Delta t]
      -
      \varphi_{k}^{(n-1)}[t_{\rm init}+(n-1)\Delta t]
    \right\}^{2}
  }.
\end{split}
\end{equation}
Then, $\mathbf{m}_{k}^{(n)}[t_{\rm init}+(n-1)\Delta t]-\mathbf{m}_{k}[t_{\rm init}+(n-1)\Delta t]$ points to the most expanded direction. 
Note that, according to the definition, $\mathbf{m}_{k}^{(n)}[t_{\rm init}+(n-1)\Delta t]$ satisfies 
\begin{equation}
\begin{split}
  \mathscr{D}^{(n)}[t_{\rm init}+(n-1)\Delta t]
  &=
  \sum_{k=1}^{N}
  \mathscr{D}_{k}^{(n)}\{\mathbf{m}_{k}[t_{\rm init}+(n-1)\Delta t],\mathbf{m}_{k}^{(n)}[t_{\rm init}+(n-1)\Delta t]\}
\\
  &=
  \sum_{k=1}^{N}
  \sqrt{
    \left\{
      \theta_{k}[t_{\rm init}+(n-1)\Delta t]
      -
      \theta_{k}^{(n)}[t_{\rm init}+(n-1)\Delta t]
    \right\}^{2}
    +
    \left\{
      \varphi_{k}[t_{\rm init}+(n-1)\Delta t]
      -
      \varphi_{k}^{(n)}[t_{\rm init}+(n-1)\Delta t]
    \right\}^{2}
  }
\\
  &=
  \varepsilon, 
\end{split}
\end{equation}
i.e., the sum of the distance between $\mathbf{m}_{k}[t_{\rm init}+(n-1)\Delta t]$ and $\mathbf{m}_{k}^{(n)}[t_{\rm init}+(n-1)\Delta t]$ is $\varepsilon$. 
Then, we evaluate the time evolution of $\mathbf{m}_{k}[t_{\rm init}+(n-1)\Delta t]$ and $\mathbf{m}_{k}^{(n)}[t_{\rm init}+(n-1)\Delta t]$ to $\mathbf{m}_{k}(t_{\rm init}+n\Delta t)$ and $\mathbf{m}_{k}(t_{\rm init}+n\Delta t)$ and define their distance as 
\begin{equation}
  \mathscr{D}_{k}^{(n)}[\mathbf{m}_{k}(t_{\rm init}+n\Delta t),\mathbf{m}_{k}^{(n)}(t_{\rm init}+n\Delta t)]
  =
  \sqrt{
    \left[
      \theta_{k}(t_{\rm init}+n\Delta t)
      -
      \theta_{k}^{(n)}(t_{\rm init}+\Delta t)
    \right]^{2}
    +
    \left[
      \varphi_{k}(t_{\rm init}+\Delta t)
      -
      \varphi_{k}^{(n)}(t_{\rm init}+\Delta t)
    \right]^{2}
  }.
\end{equation}
Using their sum, $\mathscr{D}^{(n)}(t_{\rm init}+n\Delta t)=\sum_{k=1}^{N}\mathscr{D}_{k}^{(n)}[\mathbf{m}_{k}(t_{\rm init}+n\Delta t),\mathbf{m}_{k}^{(n)}(t_{\rm init}+n\Delta t)]$, the temporal Lyapunov exponent at $t=t_{\rm init}+n\Delta t$ is defined as 
\begin{equation}
  \varLambda^{(n)}
  =
  \frac{1}{\Delta t}
  \ln
  \frac{\mathscr{D}^{(n)}(t_{\rm init}+n\Delta t)}{\varepsilon}. 
\end{equation}
Repeating these procedures, the maximum Lyapunov is obtained as a long-time average of the temporal Lyapunov exponent as 
\begin{equation}
  \varLambda
  =
  \lim_{\mathscr{N}\to \infty}
  \frac{1}{\mathscr{N}}
  \sum_{i=1}^{\mathscr{N}}
  \varLambda^{(i)}. 
\end{equation}
In this work, we start with evaluating the temporal Lyapunov exponent when the first washout inputs are injected. 
Since there are two $300$ washout inputs, $1000$ training inputs, and $1000$ test inputs, and the pulse width is $5.0$ ns, as mentioned above, the number of the averaging (or equivalently, the number of the temporal Lyapunov exponent) is $1.3\times 10^{7}$ (recall that $\Delta t=1.0$ ps). 







\section*{Acknowledgements}

The author is grateful to Takehiko Yorozu, Hikaru Nomura, Teijiro Isokawa, Hitoshi Kubota, and Yoshishige Suzuki for valuable discussion. 
This work was support by JSPS KAKENHI Grant, Number 20H05655, 24K01336 and 24K00547.


\section*{Author contributions statement}

T.T. designed the project, made code and performed the LLG simulations, prepared figures, and wrote the manuscript.  


\section*{Competing interests}

The authors declare no competing interests. 


\section*{Data availability}

The datasets used and/or analyses during the current study available from the corresponding author on reasonable request.


\section*{Additional information}

\textbf{Correspondence} and requests for materials should be addressed to T.T.





\end{document}